\documentclass[letter,11pt]{article}
\pdfoutput=1
\usepackage{jheppub}
\usepackage[dvipsnames,table]{xcolor}
\usepackage[T1]{fontenc}
\usepackage[colorlinks=true, linkcolor=blue, urlcolor=blue, citecolor=blue, anchorcolor=blue]{hyperref}
\usepackage{subcaption}
\usepackage{multirow,makecell}
\usepackage{comment}

\newcommand{\bec}{\begin{center}}
\newcommand{\eec}{\end{center}}
\newcommand{\beq}{\begin{equation}}
\newcommand{\eeq}{\end{equation}}
\newcommand{\bea}{\begin{eqnarray}}
\newcommand{\eea}{\end{eqnarray}}

\title{Probing Probability Geometry with Schwinger--Dyson Identities: Score Mismatch, Fisher Information, and Configurational Temperature}

\author[a,b]{Anosh Joseph}

\affiliation[a]{National Institute for Theoretical and Computational Sciences, \\ School of Physics, and Mandelstam Institute for Theoretical Physics,\\ University of the Witwatersrand, Johannesburg, Wits 2050, South Africa}
\affiliation[b]{Brown Center for Theoretical Physics and Innovation, \\Department of Physics, Brown University, \\Providence, RI 02912, United States}

\emailAdd{anosh.joseph@wits.ac.za}

\abstract{
We develop a geometric interpretation of Schwinger--Dyson identities by showing that their violations are controlled by a single score-mismatch field $\delta s$. For an arbitrary sampled probability distribution $Q$ and equilibrium measure $P_{\rm eq}$, every Schwinger--Dyson violation is determined by $\delta s = \nabla \log (Q / P_{\rm eq})$, which characterizes the departure from equilibrium. Each Schwinger--Dyson identity measures a projection of this field onto a probe direction in configuration space. The relative Fisher information is its squared norm. This gives a universal bound relating Fisher information to the complete Schwinger--Dyson hierarchy, thus implying that convergence in Fisher information restores all Schwinger--Dyson identities. We further obtain a variational characterization of the relative Fisher information in terms of Schwinger--Dyson violations, leading to a natural tomographic interpretation in which increasingly rich families of probe fields encode progressively more information about the underlying probability distortion. The configurational temperature, within this framework, emerges as a distinguished Schwinger--Dyson probe. The Stein operators and score-function methods arise naturally from the same probability-geometric structure. The score-mismatch field, therefore, provides a unified geometric language for understanding Schwinger--Dyson identities, configurational temperature, Fisher information, and non-equilibrium sampling in stochastic processes.
}

\begin{document}
\maketitle
\flushbottom

\section{Introduction}

Schwinger--Dyson identities occupy a central position in statistical mechanics, quantum field theory, and stochastic dynamics. Arising from the invariance of the partition function under infinitesimal changes of integration variables, they form an infinite hierarchy of exact constraints satisfied by the equilibrium probability measure. Because of their generality, Schwinger--Dyson identities play an important role in lattice field theory, matrix models, Monte Carlo simulations, stochastic quantization, and complex Langevin dynamics. Any correctly sampled equilibrium ensemble must satisfy the full hierarchy.

In practice, however, we are often interested not in the identities themselves but in their violations. Numerical simulations rarely sample the exact equilibrium distribution. 
Finite statistics introduce statistical uncertainties, while incomplete equilibration, discretization effects, approximate algorithms, and incorrect convergence produce genuine departures from the target measure. 
Consequently, Schwinger--Dyson identities are frequently used as diagnostics of sampling quality. Closely related ideas appear in configurational-temperature methods, where exact identities involving derivatives of the action are used to monitor equilibration and assess sampling accuracy.

Despite their widespread use, the meaning of Schwinger--Dyson violations remains surprisingly obscure. A nonzero violation is usually interpreted as evidence that the sampled distribution differs from the desired equilibrium measure. However, this observation raises a deeper question: what exactly is being measured? What geometric or probabilistic information is encoded in a Schwinger--Dyson violation? How should different violations be compared? What aspect of the underlying probability distribution do they probe?

The purpose of this work is to answer these questions. We develop a geometric formulation of Schwinger--Dyson identities in which violations acquire a direct interpretation in terms of probability-space geometry. 
Our starting point is the observation that Schwinger--Dyson identities arise from infinitesimal deformations generated by vector fields on configuration space and may therefore be viewed as geometric statements about probability measures. 
Within this framework, configurational temperature \cite{Dhindsa:2025xfv} emerges naturally as a distinguished member of a much larger hierarchy generated by arbitrary vector fields on configuration space.

The central result of the paper is the identification of Schwinger--Dyson violations with projections of a single geometric object, $\delta s$. For an arbitrary sampled probability distribution $Q$, and equilibrium measure $P_{\rm eq}$, we show that the violation associated with a probe field $F$ takes the form
\begin{equation}
\Delta_F(Q) = -\left\langle
F\cdot\delta s
\right\rangle_Q,
\end{equation}
where
\begin{equation}
\delta s \equiv \nabla\log\frac{Q}{P_{\rm eq}}.
\end{equation}
The vector field $\delta s$ measures the difference between the sampled and equilibrium probability geometries. We refer to it as the \emph{score mismatch} \cite{JMLR:v6:hyvarinen05a}. Schwinger--Dyson violations are therefore projections of the score mismatch onto chosen probe directions. In this interpretation, they do not merely signal the failure of an exact identity; they resolve specific directional components of the departure from equilibrium.

This observation naturally connects Schwinger--Dyson identities with information geometry. The score mismatch defines a vector field on configuration space whose squared norm is precisely the relative Fisher information,
\begin{equation}
I(Q|P_{\rm eq}) = \left\langle
|\delta s|^2
\right\rangle_Q.
\end{equation}
We show that every Schwinger--Dyson violation obeys a universal Cauchy--Schwarz bound controlled by the relative Fisher information. As a consequence, convergence in Fisher information implies the simultaneous restoration of the entire Schwinger--Dyson hierarchy. In this picture, Fisher information measures the overall magnitude (in the $L^2(Q)$ sense) of the probability distortion, while individual Schwinger--Dyson identities resolve its directional structure. This geometric interpretation is closely related to the role played by Fisher information in information geometry, where it defines a natural metric on spaces of probability distributions \cite{amari2000methods, amari2016information}.

The score-mismatch field introduced here corresponds to the difference between score functions that appear in statistics and machine learning \cite{JMLR:v6:hyvarinen05a}. 
It is also closely connected to Stein identities and Stein discrepancy constructions, where operators of the form
\begin{equation}
\nabla\cdot F + F\cdot\nabla\log P
\end{equation}
are used to quantify departures from a target distribution through families of test functions \cite{Stein1986, LiuLeeJordan2016, GorhamMackey2015, pmlr-v70-gorham17a}. The present work develops a complementary perspective rooted in Schwinger--Dyson identities, configurational temperature, and stochastic sampling diagnostics. In this framework, score mismatch, Fisher information, and Schwinger--Dyson violations emerge as different manifestations of a common probability geometry.

A central theme of this work is that Schwinger--Dyson identities may be interpreted as probes of probability geometry. Since each identity measures only a single projection of the score mismatch, different probe fields generally contain complementary information. This observation leads naturally to a tomographic viewpoint. The collection of Schwinger--Dyson violations generated by a sufficiently rich family of probe fields may be regarded as a set of measurements of an underlying score-mismatch field. Individual violations reveal directional information, while Fisher information measures the overall strength of the distortion. In this sense, the Schwinger--Dyson hierarchy admits a tomographic interpretation.

Beyond its conceptual interest, this framework has direct applications to stochastic sampling algorithms. For example, during Langevin evolution, the sampled distribution generally differs from its equilibrium form, producing a nonvanishing score mismatch and corresponding Schwinger--Dyson violations. The geometric language developed here, therefore, provides a natural description of equilibration. It suggests new diagnostics for Monte Carlo simulations and stochastic quantization. 
The geometric viewpoint developed here also suggests possible extensions to complex Langevin dynamics, where Schwinger--Dyson identities play a central role in modern correctness criteria.

More broadly, related score-function structures appear in modern score-based diffusion models and generative stochastic differential equations, where evolving score fields play a central role in the characterization of probability distributions \cite{NEURIPS2019_3001ef25, Song:2020hus, NEURIPS2020_4c5bcfec}.

The paper is organized as follows. In Sec.~\ref{sec:Schwinger--Dyson_Geometry} we develop the geometric interpretation of Schwinger--Dyson identities and show how configurational temperature emerges as a distinguished member of the hierarchy. In Sec.~\ref{sec:Non-Equilibrium_Probability_Geometry} we extend the framework away from equilibrium and introduce the score mismatch field. Section~\ref{sec:Fisher_Information_and_SD_Violations} establishes Fisher-information bounds on Schwinger--Dyson violations and explores their information-geometric interpretation. Applications to stochastic quantization and sampling diagnostics are discussed in Sec.~\ref{sec:Applications_and_Diagnostics}. In Sec.~\ref{sec:Tomography_of_Probability_Geometry} we develop the tomographic interpretation of Schwinger--Dyson violations and investigate the information contained in the hierarchy as a whole. We conclude in Sec.~\ref{sec:Discussion_and_Outlook} with a discussion of future directions and potential applications.

\section{Schwinger--Dyson Geometry}
\label{sec:Schwinger--Dyson_Geometry}

\subsection{General Schwinger--Dyson Identities}

Consider a configuration space ${\cal C}$, which is an infinite-dimensional manifold, whose points are field configurations $\phi(x)$. 
Tangent vectors at a point $\phi$ are infinitesimal field variations $\delta \phi(x)$. 
The starting point of our discussion is the Euclidean path integral
\begin{equation}
Z = \int {\cal D}\phi \, e^{-S_E[\phi]},
\label{eq:partition_function}
\end{equation}
where $S_E[\phi]$ is the Euclidean action. 
Throughout this work we adopt the viewpoint that the Boltzmann weight
\begin{equation}
P_{\rm eq}[\phi] = \frac{1}{Z} e^{-S_E[\phi]}
\label{eq:equilibrium_measure}
\end{equation}
defines a probability distribution on the infinite-dimensional configuration space of fields. 

From this perspective, the path integral may be viewed as an expectation value with respect to the equilibrium probability density $P_{\rm eq}$. 
The Euclidean action determines how probability is distributed across configuration space, with configurations of smaller action receiving larger statistical weight. 

The fundamental observation underlying Schwinger--Dyson identities is that the partition function is invariant under a change of integration variables. 
Consider an infinitesimal field redefinition
\begin{equation} 
\phi(x) \rightarrow \phi(x) + \epsilon \, F[\phi],
\label{eq:field_redefinition}
\end{equation}
where $F[\phi]$ is an arbitrary functional vector field and $\epsilon$ is infinitesimal. 
Geometrically, $F[\phi]$ is a tangent vector field on the infinite-dimensional configuration space. 
The transformation (\ref{eq:field_redefinition}) moves every configuration infinitesimally along this vector field.

Under this transformation the functional measure formally changes according to
\begin{equation}
{\cal D} \phi' = {\cal D} \phi \left( 1 + \epsilon \, \nabla \cdot F \right),
\label{eq:measure_change}
\end{equation}
where
\begin{equation}
\nabla \cdot F = \int d^dx \, \frac{\delta F(x)} {\delta\phi(x)}
\label{eq:functional_div}
\end{equation}
is the functional divergence of the vector field.

Similarly, the action varies as
\begin{equation}
S_E[\phi + \epsilon F] = S_E[\phi] + \epsilon \, F \cdot \nabla S_E + {\cal O}(\epsilon^2),
\label{eq:action_variation_jhep}
\end{equation}
with
\begin{equation}
F \cdot \nabla S_E = \int d^dx \, F(x) \frac{\delta S_E}{\delta\phi(x)}.
\end{equation}

Since Eq.~(\ref{eq:field_redefinition}) is merely a change of variables, the partition function must remain unchanged,
\begin{equation}
Z = \int {\cal D} \phi' \, e^{-S_E[\phi']} = \int {\cal D} \phi \, e^{-S_E[\phi]}.
\end{equation}

Expanding to first order in $\epsilon$, and assuming that the boundary terms vanish, gives
\begin{equation}
0 = \int {\cal D} \phi \, e^{-S_E[\phi]} \left( \nabla \cdot F - F \cdot \nabla S_E \right).
\end{equation}

Dividing by the partition function yields
\begin{equation}
\left\langle \nabla \cdot F \right\rangle_{\rm eq} = \left\langle F \cdot \nabla S_E \right\rangle_{\rm eq},
\label{eq:SD_master}
\end{equation}
which is the general Schwinger--Dyson identity associated with the vector field $F$.

Equation~(\ref{eq:SD_master}) is one of the most general exact relations satisfied by the equilibrium measure. 
Conventional Schwinger--Dyson equations are recovered by choosing $F$ to be local functions of the fields. 
The collection of all such equations forms an infinite hierarchy of constraints on the probability distribution.

It is worth pausing to interpret Eq.~(\ref{eq:SD_master}) geometrically. 
The vector field $F$ specifies an infinitesimal deformation of configuration space. 
The quantity $\nabla \cdot F$ measures the local change in volume associated with this deformation, while $F \cdot \nabla S_E$ measures the corresponding change in the Boltzmann weight. 
The Schwinger--Dyson identity states that these two effects cancel exactly when averaged over the equilibrium measure. 

In this sense, Schwinger--Dyson identities are not isolated algebraic relations but rather consequences of a geometric balance between volume deformation and probability flow in configuration space. 
This geometric viewpoint will serve as the foundation for the developments that follow.

\subsection{Configurational Temperature as a Distinguished Schwinger--Dyson Identity}

Equation~(\ref{eq:SD_master}) generates an infinite family of exact identities, one for each choice of the vector field $F$. 
Most choices are equally valid from a mathematical perspective, but some are more natural than others from a physical point of view. 

A particularly useful choice of $F$ is obtained by asking whether one can construct a vector field whose contraction with the action gradient is identically unity, 
\begin{equation} 
F \cdot \nabla S_E = 1. 
\label{eq:unit_projection} 
\end{equation} 
Such a vector field probes the local direction of steepest increase of the action while rescaling the gradient so that its contraction with $\nabla S_E$ is identically unity. 

A natural choice satisfying Eq.~(\ref{eq:unit_projection}) is a gradient flow, that has the form,
\begin{equation}
F = v = \frac{\nabla S_E}{|\nabla S_E|^2}.
\label{eq:normalized_gradient_flow} 
\end{equation} 
Substituting Eq.~(\ref{eq:normalized_gradient_flow}) into the general Schwinger--Dyson identity (\ref{eq:SD_master}) immediately gives 
\begin{equation} 
\left\langle \nabla \cdot v \right\rangle_{\rm eq} = 1, 
\label{eq:config_temp_identity} 
\end{equation} 
or equivalently 
\begin{equation} 
\left\langle \nabla \cdot \left( \frac{\nabla S_E} {|\nabla S_E|^2} \right) \right\rangle_{\rm eq} = 1. 
\end{equation} 
Equation~(\ref{eq:config_temp_identity}) is the configurational-temperature identity \cite{Dhindsa:2025xfv, Joseph:2025fcd, Joseph:2025xbn, Longia:2026doi, Joseph:2026xti}. 
Unlike conventional thermodynamic definitions of temperature, it depends only on derivatives of the action and therefore involves purely configurational information.

Historically, configurational temperature was introduced in classical statistical mechanics as an alternative estimator of inverse temperature constructed directly from the geometry of the energy landscape \cite{Rugh:1997, Rugh:1998, Butler:1998, Baranyai:2000, Rickayzen:2001, Palma:2016, Jepps:2000}. 
From the present perspective, however, Eq.~(\ref{eq:config_temp_identity}) is not an isolated relation. 
Rather, it appears as a distinguished member of the Schwinger--Dyson hierarchy associated with the gradient flow (\ref{eq:normalized_gradient_flow}). 

It is therefore natural to define the configurational-temperature observable 
\begin{equation} 
\beta_{\rm config} \equiv \left\langle \nabla \cdot \left( \frac{\nabla S_E} {|\nabla S_E|^2} \right) \right\rangle. 
\label{eq:beta_config} 
\end{equation} 
At equilibrium, one has 
\begin{equation}
\beta_{\rm config} = 1, 
\end{equation} 
while deviations from unity signal departures from the equilibrium measure. 

The significance of Eq.~(\ref{eq:beta_config}) extends beyond this particular observable. 
Since it arises from a specific choice of probe field, configurational temperature inherits properties shared by the entire Schwinger--Dyson hierarchy. 
Understanding these common geometric structures will be one of the main goals of this work. 

\subsection{Schwinger--Dyson Identities in Terms of the Score Field} 

The derivation above suggests that Schwinger--Dyson identities are fundamentally geometric statements about probability measures on configuration space. This interpretation becomes more transparent when Eq.~(\ref{eq:SD_master}) is rewritten in terms of the equilibrium probability distribution itself.

From the equilibrium measure $P_{\rm eq}[\phi]$ we define the corresponding score field 
\begin{equation} 
s_{\rm eq} \equiv \nabla \log P_{\rm eq}. 
\label{eq:score_definition} 
\end{equation} 
The score field describes the local variation of the probability density across configuration space. 
Regions in which the probability density changes rapidly correspond to large score fields, while regions of nearly uniform probability correspond to small score fields. 

Using the Boltzmann form of the equilibrium measure, $P_{\rm eq} \propto e^{-S_E}$, one immediately finds 
\begin{equation} 
s_{\rm eq} = -\nabla S_E. 
\label{eq:score_action_relation} 
\end{equation}

Substituting Eq.~(\ref{eq:score_action_relation}) into the Schwinger--Dyson identity (\ref{eq:SD_master}) yields 
\begin{equation} 
\left\langle \nabla \cdot F + F \cdot s_{\rm eq} \right\rangle_{\rm eq} = 0. 
\label{eq:sd_score_form} 
\end{equation} 

This form of the identity is particularly revealing. 
The explicit dependence on the action is now replaced by the score field $s_{\rm eq}$, which encodes how the probability density changes across configuration space. 

The quantity
\begin{equation}
\mathcal A_F \equiv \nabla \cdot F + F \cdot s_{\rm eq} = \frac{1}{P_{\rm eq}} \nabla \cdot (P_{\rm eq}F)
\label{eq:stein_operator}
\end{equation}
admits a natural geometric interpretation. 
The vector field $F$ generates an infinitesimal deformation of configuration space, while $P_{\rm eq}F$ may be viewed as the corresponding probability current. 
The operator $\mathcal A_F$ is therefore the weighted divergence of this probability current with respect to the equilibrium measure. 

Equation~(\ref{eq:sd_score_form}) states that the equilibrium measure is invariant under such infinitesimal deformations when averaged over the ensemble,
\begin{equation}
\left\langle \mathcal A_F \right\rangle_{\rm eq} = \left\langle \frac{1}{P_{\rm eq}} \nabla\cdot(P_{\rm eq}F) \right\rangle_{\rm eq} 
= \int D\phi \, \nabla\cdot(P_{\rm eq}F) = 0.
\end{equation} 
This identity follows simply from the vanishing of the total divergence under the usual assumption that boundary contributions vanish. 

It is worth noting that the operator $\mathcal A_F$ is precisely the Stein operator associated with the target probability distribution $P_{\rm eq}$, which plays a central role in Stein's method and Stein discrepancy constructions \cite{Stein:1972, Stein1986, GorhamMackey2015}. 
From this perspective, the Schwinger--Dyson identities may be viewed as the statement that the expectation value of the Stein operator vanishes with respect to the equilibrium measure. 
This connection provides a natural bridge between Schwinger--Dyson identities in field theory and modern score-based methods in probability, statistics, and machine learning. 

The score field, therefore, emerges as the central geometric object underlying the Schwinger--Dyson hierarchy. 
Rather than viewing Schwinger--Dyson identities as relations involving derivatives of the action, one may regard them as constraints determined by the geometry of the probability measure itself. 

This reformulation is particularly useful away from equilibrium. 
If the sampled probability distribution differs from $P_{\rm eq}$, the corresponding score field also changes. 
The resulting mismatch between the score fields will be precisely the quantity measured by Schwinger--Dyson violations. Developing this idea will be the focus of the next section. 

\section{Non-Equilibrium Probability Geometry} 
\label{sec:Non-Equilibrium_Probability_Geometry}

The discussion so far has been restricted to the equilibrium measure $P_{\rm eq}$. 
For this distribution, the Schwinger--Dyson hierarchy is satisfied exactly, reflecting the invariance of the path integral under infinitesimal deformations of configuration space.

In many situations of physical interest, however, one encounters probability distributions that differ from the equilibrium measure. 
Examples include Monte Carlo simulations before equilibration, finite Langevin-time distributions in stochastic quantization, and stationary distributions generated by approximate sampling algorithms. 
In such cases, the Schwinger--Dyson identities need no longer vanish. 

This immediately raises a natural question: if Schwinger--Dyson identities are violated, what exactly are they measuring? 

The answer is surprisingly simple. 
We shall show that Schwinger--Dyson violations are controlled by the score mismatch $\delta s$.

\subsection{Generalized Schwinger--Dyson Violations} 

Let $Q[\phi]$ be an arbitrary normalized probability distribution on configuration space. 
We assume throughout that $Q$ is sufficiently smooth, strictly positive on its support, and satisfies the same boundary conditions required for the integration by parts arguments below. 

Even when $Q$ is not an equilibrium measure, one may still evaluate the Schwinger--Dyson operator associated with a vector field $F$, 
\begin{equation} 
\Delta_F(Q) = \left\langle \nabla \cdot F - F \cdot \nabla S_E \right\rangle_Q. 
\label{eq:deltaF_definition} 
\end{equation} 

The quantity $\Delta_F(Q)$ measures the extent to which the Schwinger--Dyson identity associated with $F$ fails to hold in the distribution $Q$. 

At equilibrium, $Q = P_{\rm eq}$, the Schwinger--Dyson identity implies 
\begin{equation} 
\Delta_F(P_{\rm eq}) = 0 
\end{equation} 
for every admissible vector field $F$. 
Away from equilibrium, however, the quantity (\ref{eq:deltaF_definition}) is generally nonzero. 

At first sight, the expression (\ref{eq:deltaF_definition}) appears to depend separately on the probe field $F$, the action $S_E$, and the probability distribution $Q$. 
It is therefore not obvious whether the resulting violation possesses any simple geometric interpretation. 

To answer this question, let us rewrite Eq.~(\ref{eq:deltaF_definition}) in a form that depends directly on the probability distribution itself. 

Using integration by parts with respect to the measure $Q$ gives 
\begin{equation} 
\left\langle \nabla \cdot F \right\rangle_Q = - \left\langle F \cdot \nabla \log Q \right\rangle_Q, 
\label{eq:integration_by_parts_Q} 
\end{equation} 
assuming that boundary contributions vanish. 

Substituting Eq.~(\ref{eq:integration_by_parts_Q}) into Eq.~(\ref{eq:deltaF_definition}) yields 
\begin{equation} 
\Delta_F(Q) = - \left\langle F \cdot \left( \nabla \log Q + \nabla S_E \right) \right\rangle_Q. 
\label{eq:delta_intermediate} 
\end{equation} 

At this stage, an important simplification occurs. 
In the previous section, we introduced the equilibrium score field $s_{\rm eq} = - \nabla S_E = \nabla \log P_{\rm eq}$. 
Using this relation, Eq.~(\ref{eq:delta_intermediate}) becomes 
\begin{equation} 
\Delta_F(Q) = \left\langle F \cdot \left( \nabla \log Q - \nabla \log P_{\rm eq} \right) \right\rangle_Q. 
\end{equation} 

Combining the logarithms gives the remarkably compact expression
\begin{equation}
\Delta_F(Q) = - \left\langle F \cdot \nabla \log \frac{Q}{P_{\rm eq}} \right\rangle_Q.
\label{eq:DeltaScore}
\end{equation} 

It shows that the Schwinger--Dyson violation can be expressed entirely in terms of the sampled probability distribution $Q$ and the target equilibrium measure $P_{\rm eq}$, without explicit reference to the action or its derivatives. 
The relevant geometric quantity is the logarithmic derivative of the probability ratio,
\begin{equation}
\nabla \log \frac{Q}{P_{\rm eq}},
\label{eq:prob_ratio}
\end{equation}
which measures the difference between the sampled and equilibrium probability geometries. 

Equation~(\ref{eq:DeltaScore}) therefore reveals that every Schwinger--Dyson violation is probing the same underlying geometric object. 
Different choices of the probe field $F$ do not measure different departures from equilibrium; rather, they resolve different directional components of a single probability-space distortion. 
This observation forms the basis for the geometric interpretation developed throughout the remainder of the paper. 

\subsection{The Score-Mismatch Field} 

Equation~(\ref{eq:DeltaScore}) suggests that the geometric content of Schwinger--Dyson violations is encoded in the quantity Eq. \eqref{eq:prob_ratio}. 
To understand its significance, let us look at it using the score field notation.

The equilibrium measure defines an equilibrium score field $s_{\rm eq}$. 
For a normalized probability distribution $Q$, we have the score field 
\begin{equation} 
s_Q = \nabla \log Q. 
\label{eq:score_Q} 
\end{equation} 

Using these definitions, we can introduce a vector field
\begin{equation} 
\delta s = s_Q - s_{\rm eq}.
\label{eq:delta_s_definition} 
\end{equation} 

It measures the difference between the sampled and equilibrium score fields. 
Explicitly, 
\begin{equation} 
\delta s = \nabla \log \frac{Q}{P_{\rm eq}}. 
\label{eq:ScoreMismatch} 
\end{equation} 

We shall refer to $\delta s$ as the \emph{score mismatch}. 
By construction, $\delta s = 0$ if and only if $Q = P_{\rm eq}$, so the score mismatch provides a local measure of departure from equilibrium. 

We see that Eq.~(\ref{eq:DeltaScore}) may be rewritten in the remarkably simple form
\begin{equation}
\Delta_F(Q) = - \left\langle F \cdot \delta s \right\rangle_Q.
\label{eq:DeltaProjection}
\end{equation}

Equation~(\ref{eq:DeltaProjection}) is the central result of this work. 
It reveals the geometric structure underlying the entire Schwinger--Dyson hierarchy. Every admissible probe field $F$ defines a Schwinger--Dyson identity and therefore a corresponding violation $\Delta_F(Q)$. Since there are infinitely many possible choices of $F$, the Schwinger--Dyson hierarchy may be viewed as an infinite family of probes on configuration space.

The significance of Eq.~(\ref{eq:DeltaProjection}) is that all of these probes are measuring the same underlying geometric object, namely the score mismatch $\delta s$. 
No matter which probe field is chosen, the corresponding Schwinger--Dyson violation is determined entirely by its projection onto this single vector field. The infinite Schwinger--Dyson hierarchy, therefore, does not probe different departures from equilibrium; rather, it resolves different directional components of one underlying probability-space distortion.

Equation~(\ref{eq:DeltaProjection}) consequently admits a simple geometric interpretation. The score mismatch $\delta s$ characterizes the local deformation of the sampled probability geometry relative to equilibrium, while the probe field $F$ selects a direction in configuration space. The Schwinger--Dyson violation $\Delta_F(Q)$ is the projection of the score mismatch onto that direction and may therefore be viewed as a directional measurement of the underlying probability distortion.

This observation leads to a useful reinterpretation of Schwinger--Dyson identities. 
Traditionally, they are regarded as exact consistency conditions satisfied by the equilibrium measure. 
From the present perspective, however, they may also be viewed as probes of probability geometry. 
When the sampled distribution differs from equilibrium, different Schwinger--Dyson identities respond to different components of the score mismatch. 

The situation is analogous to measuring an ordinary vector field. 
A single projection reveals only one component of the vector, whereas measurements along many directions reveal increasingly detailed information about its structure. 
Likewise, a single Schwinger--Dyson identity probes only one component of the score mismatch, whereas the full hierarchy encodes the entire probability distortion. 

The score mismatch $\delta s$ plays a central role in score-based statistical inference and machine learning. 
In particular, the integration-by-parts identity used in Eq. \eqref{eq:integration_by_parts_Q} is the same mathematical step that underlies score matching, where it is employed to rewrite the Fisher divergence between data and model distributions in a computable form~\cite{JMLR:v6:hyvarinen05a}. 
The present work provides a complementary interpretation of this object in the context of Euclidean field theory, where it emerges naturally from the geometric structure of the Schwinger--Dyson identities.

This geometric interpretation will play a central role in what follows. 
In particular, it suggests that Schwinger--Dyson violations should not merely be regarded as indicators of incorrect sampling. 
Rather, they contain quantitative information about how the sampled distribution differs from the equilibrium distribution. 
The next sections will make this statement precise by relating the score mismatch to Fisher information and by showing how families of Schwinger--Dyson identities can be used to probe the geometry of sampling errors. 

\subsection{Example: A Shifted Gaussian Distribution}

Before proceeding further, it is useful to examine a simple example in which the score mismatch can be computed explicitly. 
Although elementary, this example already illustrates the central geometric ideas developed above. 

Consider the equilibrium probability distribution 
\begin{equation} 
P_{\rm eq}(x) = \frac{1}{\sqrt{2 \pi}} e^{-x^2/2}, 
\label{eq:gaussian_equilibrium} 
\end{equation} 
which corresponds to a unit-variance Gaussian centered at the origin. 
Suppose that the sampled distribution is instead 
\begin{equation} 
Q(x) = \frac{1}{\sqrt{2\pi}} e^{-(x - \mu)^2/2}, 
\label{eq:shifted_gaussian} 
\end{equation} 
where $\mu$ is a constant displacement parameter. 

The two probability distributions differ only by a translation of their mean. One may therefore expect the corresponding probability geometries to differ in a similarly simple manner. 

The equilibrium score field is $s_{\rm eq} = - x$, while the score field associated with the shifted distribution is $s_Q = - (x - \mu)$. 
The score mismatch therefore, becomes
\begin{equation} 
\delta s = \mu. 
\label{eq:gaussian_score_mismatch} 
\end{equation} 

Several features of this result are worth noting. 

First, the score mismatch is independent of position. 
Although the probability distributions themselves vary across configuration space, their difference is encoded in a constant field. 
A uniform displacement of the score field represents the entire departure from equilibrium. 

Second, the magnitude of the score mismatch is controlled directly by the parameter $\mu$. 
As the two distributions approach one another, the score mismatch decreases continuously and vanishes when $\mu = 0$. 

Substituting Eq.~(\ref{eq:gaussian_score_mismatch}) into the general expression (\ref{eq:DeltaProjection}) yields 
\begin{equation} 
\Delta_F(Q) = - \mu \langle F\rangle_Q. 
\label{eq:gaussian_violation} 
\end{equation} 

The Schwinger--Dyson violation is therefore proportional to the displacement parameter and depends on the probe field only through its expectation value. 
Different probes do not detect different distortions; rather, they measure the same underlying score mismatch with different sensitivities. 

This example provides a simple illustration of the general picture developed in the previous subsection. 
The Schwinger--Dyson violations do not directly measure differences between probability distributions. 
Instead, they measure projections of the score-mismatch field that relate the distributions. In the present case, the score mismatch is particularly simple, consisting of a constant translation. 
More complicated probability distortions lead to nontrivial score-mismatch fields, whose structure can be resolved by suitable choices of probe field. 

\subsection{Example: A Quartic Deformation} 

To appreciate the richer possibilities that arise in more general situations, it is instructive to consider a deformation that changes the shape of the probability distribution rather than merely translating it. 

Let us use the same equilibrium distribution $P_{\rm eq}(x)$ as before, and now consider the deformed probability density 
\begin{equation} 
Q(x) = \frac{1}{Z_Q} \exp \left( - \frac{x^2}{2} - \lambda x^4 \right), 
\label{eq:quartic_Q} 
\end{equation} 
where $\lambda > 0$ controls the strength of the deformation. 

Unlike the shifted Gaussian, this modification does not change the location of the maximum of the distribution. 
Instead, it suppresses configurations with large $|x|$, yielding a probability density that is more sharply localized around the origin. 

The corresponding score fields are: $s_{\rm eq} = - x$ and $s_Q = - x - 4 \lambda x^3$. 
The score mismatch therefore, takes the form 
\begin{equation} 
\delta s = s_Q - s_{\rm eq} = - 4 \lambda x^3. 
\label{eq:quartic_score} 
\end{equation} 

Several important differences from the previous example are immediately apparent. 

First, the score mismatch is no longer constant. 
Instead, it varies across configuration space and becomes increasingly large away from the origin. 
This behavior reflects the fact that the quartic deformation acts most strongly on large-field configurations, where the probability density differs most significantly from the equilibrium Gaussian. 

Second, the score mismatch now contains information about the detailed structure of the probability distortion. 
The departure from equilibrium is no longer characterized by a single parameter but by a nontrivial position-dependent vector field. 

Substituting Eq.~(\ref{eq:quartic_score}) into the general expression (\ref{eq:DeltaProjection}) gives 
\begin{equation} 
\Delta_F(Q) = - 4 \lambda \left\langle F(x) \, x^3 \right\rangle_Q. 
\label{eq:quartic_violation} 
\end{equation} 

The Schwinger--Dyson violation, therefore, depends not only on the strength of the deformation but also on the choice of probe field. 
Different probes become sensitive to different aspects of the underlying score mismatch. 

For example, choosing $F(x) = 1$ gives 
\begin{equation} 
\Delta_F(Q) = - 4 \lambda \langle x^3 \rangle_Q, 
\end{equation} 
while $F(x) = x$ yields 
\begin{equation}
\Delta_F(Q) = - 4 \lambda \langle x^4 \rangle_Q. 
\end{equation} 

The two probes respond to different moments of the same probability distortion. 
More generally, different members of the Schwinger--Dyson hierarchy resolve distinct components of the score-mismatch field.
Since the deformed distribution is even under $x \to - x $, one has $\langle x^3 \rangle_Q = 0$. The probe $F = 1$ is therefore insensitive to the quartic distortion, whereas $F = x$ detects it through the non-vanishing fourth moment. This illustrates how different probes can possess markedly different sensitivities to the same underlying score mismatch.

This example highlights an important conceptual point. 
Schwinger--Dyson violations are not merely indicators that a probability distribution differs from equilibrium. 
They also contain information about the structure of that difference. 
The score mismatch acts as a geometric representation of the underlying distortion, while the various Schwinger--Dyson identities probe different aspects of that representation. 

This observation will play a central role in the remainder of the paper. 
Once the score mismatch is viewed as a geometric field on configuration space, it becomes natural to ask how to quantify its overall magnitude and how much of its structure can be inferred from a collection of Schwinger--Dyson measurements. 
These questions lead naturally to Fisher information and, ultimately, to a tomographic interpretation of the Schwinger--Dyson hierarchy. 

\section{Fisher Information and Schwinger--Dyson Violations} 
\label{sec:Fisher_Information_and_SD_Violations}

The central lesson of the previous section is that Schwinger--Dyson violations are not arbitrary quantities. 
Rather, every violation is determined by the same underlying geometric object, namely the score mismatch, $\delta s$.
Different members of the Schwinger--Dyson hierarchy probe different components of this vector field, much as different projections probe different components of an ordinary vector. 
The examples considered above illustrate that the score mismatch can possess nontrivial structure and may vary significantly across configuration space. 

This observation naturally raises a new question. While Schwinger--Dyson identities provide directional information about the score mismatch, how should one quantify its overall magnitude? 
In other words, is there a natural measure of the total departure from equilibrium that underlies the entire Schwinger--Dyson hierarchy? 

The relative Fisher information provides the answer. 
As we shall see, Fisher information plays a distinguished role because it quantifies the norm of the score-mismatch field and thereby controls the magnitude of every Schwinger--Dyson violation. 

\subsection{Fisher-Information Bound} 

The relative Fisher information of a probability distribution $Q$ with respect to the equilibrium measure $P_{\rm eq}$ is defined by
\begin{equation}
I(Q|P_{\rm eq}) = \left\langle \left| \nabla \log \frac{Q}{P_{\rm eq}} \right|^2 \right\rangle_Q. 
\label{eq:FisherDef} 
\end{equation} 

Comparing Eq.~(\ref{eq:FisherDef}) with the definition of the score mismatch immediately yields 
\begin{equation} 
I(Q|P_{\rm eq}) = \langle |\delta s|^2 \rangle_Q. 
\label{eq:FisherScore} 
\end{equation} 

This relation gives a simple geometric interpretation of Fisher information. 
The score mismatch field describes the distortion of the probability geometry, while the Fisher information measures the squared norm of that distortion averaged over the sampled distribution. 

This viewpoint is closely related to the role played by Fisher information in information geometry, where it defines a natural metric on statistical manifolds and quantifies the distinguishability of nearby probability distributions \cite{amari2000methods, amari2016information}. 
Equation~(\ref{eq:FisherScore}) shows that, within the present framework, the same quantity arises as the squared norm of the score-mismatch field.

In particular, 
\begin{equation} 
I(Q|P_{\rm eq}) \ge 0, 
\end{equation} 
with equality only when $Q = P_{\rm eq}$. 
The relative Fisher information, therefore, provides a global measure of the departure from equilibrium. 

The connection with Schwinger--Dyson violations now follows immediately. 
In the previous section, we found that, from Eq. \eqref{eq:DeltaProjection}, $\Delta_F(Q) = - \langle F \cdot \delta s \rangle_Q$. 
This expression has the form of an inner product between the probe field $F$ and the score mismatch $\delta s$. 
Applying the Cauchy--Schwarz inequality therefore, gives 
\begin{equation} 
|\Delta_F(Q)|^2 \le \langle |F|^2 \rangle_Q \, \langle |\delta s|^2\rangle_Q. 
\end{equation} 

Using Eq.~(\ref{eq:FisherScore}) we obtain 
\begin{equation} 
|\Delta_F(Q)|^2 \le \langle |F|^2 \rangle_Q \, I(Q|P_{\rm eq}). 
\label{eq:FisherBound} 
\end{equation} 

Equation~(\ref{eq:FisherBound}) is one of the principal results of this work. 

Its significance is best understood geometrically. 
The Schwinger--Dyson violation associated with a given probe field measures a particular projection of the score mismatch. In contrast, the Fisher information quantifies the total squared length of the score-mismatch field. 
Equation~(\ref{eq:FisherBound}) therefore states that no individual projection can exceed the norm of the vector field from which it originates. 

The analogy with ordinary Euclidean geometry is instructive. 
Given a vector $v$, the magnitude of its projection onto any direction is bounded by the length of the vector itself. 
Here, the score mismatch plays the role of $v$, while the Schwinger--Dyson violation corresponds to its projection along the probe field $F$. 

An immediate consequence of Eq.~(\ref{eq:FisherBound}) is that convergence in Fisher information implies the restoration of the entire Schwinger--Dyson hierarchy. Indeed, if
\begin{equation}
I(Q|P_{\rm eq}) \rightarrow 0,
\end{equation}
then
\begin{equation}
\Delta_F(Q) \rightarrow 0
\end{equation}
for every admissible probe field $F$. Thus, the vanishing of the relative Fisher information guarantees that every Schwinger--Dyson identity is simultaneously recovered. From the geometric perspective developed here, this reflects that the score mismatch field has vanishing norm, so all its projections must also vanish.

Thus, Fisher information controls all Schwinger--Dyson violations simultaneously. 
Individual Schwinger--Dyson identities provide directional information about the departure from equilibrium, whereas the Fisher information measures the overall magnitude of that departure. 
Together, they furnish complementary descriptions of the same underlying probability geometry. 

\subsection{Restoration of the Schwinger--Dyson Hierarchy} 

The Fisher-information bound (\ref{eq:FisherBound}) has an important consequence for the approach to equilibrium. 
While individual Schwinger--Dyson violations probe particular components of the score mismatch, the Fisher information controls the size of the mismatch field as a whole. 
As a result, convergence in Fisher information implies the simultaneous restoration of all Schwinger--Dyson identities. 

To see this, consider a family of probability distributions $Q_t$, labeled by a parameter $t$. 
In applications, $t$ may represent Monte Carlo time, Langevin time, or any other parameter describing the evolution of a sampling process. 
Suppose that the distributions approach equilibrium in the sense that 
\begin{equation} 
I(Q_t | P_{\rm eq}) \rightarrow 0 \qquad (t \rightarrow \infty). 
\label{eq:fisher_convergence} 
\end{equation} 

Then Eq.~(\ref{eq:FisherBound}) immediately implies 
\begin{equation} 
|\Delta_F(Q_t)|^2 \le \langle |F|^2 \rangle_{Q_t} \, I(Q_t | P_{\rm eq}). 
\end{equation} 

Provided the probe field satisfies 
\begin{equation} 
\langle |F|^2\rangle_{Q_t} < \infty, 
\end{equation} 
it follows that 
\begin{equation} 
\Delta_F(Q_t) \rightarrow 0. 
\end{equation} 

Since the choice of $F$ is arbitrary, every member of the Schwinger--Dyson hierarchy is restored in the limit of vanishing Fisher information. 
We therefore arrive at the implication 
\begin{equation} 
I(Q_t | P_{\rm eq}) \rightarrow 0 \qquad \Longrightarrow \qquad \Delta_F(Q_t) \rightarrow 0 \quad \forall \, F. 
\label{eq:hierarchy_restoration} 
\end{equation} 

This result has a natural geometric interpretation. 
The score mismatch field 
\begin{equation} 
\delta s_t = \nabla \log \frac{Q_t}{P_{\rm eq}} 
\end{equation} 
encodes the deviation of the sampled probability geometry from the equilibrium geometry at $t$. 
The Fisher information measures the squared norm of this field, 
\begin{equation} 
I(Q_t | P_{\rm eq}) = \langle |\delta s_t|^2 \rangle_{Q_t}. 
\end{equation} 

As the Fisher information decreases, the score mismatch becomes progressively smaller throughout configuration space. 
Since every Schwinger--Dyson violation is a projection of the score mismatch, 
\begin{equation} 
\Delta_F(Q_t) = - \langle F \cdot \delta s_t \rangle_{Q_t}, 
\end{equation} 
all such projections must vanish when the mismatch itself disappears. 

From this perspective, the restoration of the Schwinger--Dyson hierarchy is not a collection of independent statements. 
Rather, it is the manifestation of a single geometric phenomenon: the vanishing of the score mismatch field. 

This observation is particularly relevant in stochastic sampling algorithms. 
During equilibration, one typically monitors a small number of observables in order to assess convergence. 
Equation~(\ref{eq:hierarchy_restoration}) suggests a complementary viewpoint. 
Instead of focusing on individual observables, one may regard equilibration as the progressive restoration of the entire Schwinger--Dyson hierarchy. 
The Fisher information then provides a global measure of this process, while the individual Schwinger--Dyson violations reveal how convergence proceeds in different directions of configuration space. 

The converse question is also interesting. 
If a sufficiently large collection of Schwinger--Dyson violations becomes small, to what extent can one infer that the Fisher information itself is small? 
The vanishing of the Schwinger--Dyson violation for a single probe field, or even for a finite collection of probe fields, does not in general imply that the Fisher information is small. 
Such measurements constrain only the projections of the score mismatch onto the chosen probes, leaving components orthogonal to those probes undetermined. 
As we shall see in Sec.~\ref{sec:Tomography_of_Probability_Geometry}, it is only when one considers the complete family of admissible probe fields that the full magnitude of the score mismatch, and hence the Fisher information, can be recovered. 

\subsection{Example: Temperature Mismatch} 

The Fisher-information bound derived above admits a particularly transparent illustration in a Gaussian example. 
Unlike the shifted Gaussian discussed in the previous section, the present example changes the width of the distribution rather than its mean. 
The resulting distortion may be interpreted as a mismatch in the effective temperature of the ensemble. 

Consider the equilibrium distribution $P_{\rm eq}(x)$ and the one-parameter family of distributions 
\begin{equation} 
Q(x) = \sqrt{\frac{a}{2\pi}} e^{-a x^2 / 2}, 
\label{eq:temp_Q} 
\end{equation} 
where the parameter $a > 0$ may be viewed as an effective inverse temperature. 
For $a = 1$ the two distributions coincide, while values of $a$ different from unity correspond to a change in the width of the Gaussian. 

The corresponding score fields are: $s_{\rm eq} = - x$ and $s_Q = - a x$. 
The score mismatch, therefore, takes the simple form
\begin{equation} 
\delta s = (1 - a) x. 
\label{eq:temp_delta_s} 
\end{equation} 

Unlike the previously discussed shifted Gaussian, the mismatch is no longer constant. 
Instead, it grows linearly with distance from the origin. 
Configurations near the center of the distribution are only weakly affected, whereas configurations in the tails experience a larger distortion. 
The geometry of the probability mismatch, therefore, reflects the change in the width of the distribution. 

The corresponding Fisher information is 
\begin{equation} 
I(Q | P_{\rm eq}) = (1 - a)^2 \langle x^2 \rangle_Q. 
\end{equation} 

Using 
\begin{equation} 
\langle x^2 \rangle_Q = \frac{1}{a}, 
\end{equation} 
one finds 
\begin{equation}
I(Q | P_{\rm eq}) = \frac{(1 - a)^2}{a}. 
\label{eq:temp_fisher} 
\end{equation} 

The Fisher information, therefore, increases as the width of the sampled distribution deviates from its equilibrium value. 

To examine the Schwinger--Dyson violation, consider the probe field $F(x) = x$. 
This yields 
\begin{equation} 
\Delta_F(Q) = - \langle x \, \delta s \rangle_Q = - \frac{1 - a}{a}. 
\label{eq:temp_SD} 
\end{equation} 

It follows that 
\begin{equation}
|\Delta_F(Q)|^2 = \frac{(1 - a)^2}{a^2}, 
\end{equation} 
while 
\begin{equation} 
\langle F^2 \rangle_Q = \frac{1}{a}. 
\end{equation} 

Substituting these expressions into the Fisher-information bound (\ref{eq:FisherBound}) gives 
\begin{equation} 
\frac{(1 - a)^2}{a^2} \le \frac{1}{a} \, \frac{(1 - a)^2}{a}, 
\end{equation} 
which is satisfied as an equality. 

The reason for this exact saturation is instructive. 
From Eq.~(\ref{eq:temp_delta_s}) we see that the score mismatch is proportional to $x$, and therefore proportional to the chosen probe field, 
\begin{equation} 
\delta s = (1 - a)F. 
\end{equation} 

The probe is therefore perfectly aligned with the probability distortion. 
The Schwinger--Dyson violation measures the entire score mismatch rather than merely one of its components. 

Geometrically, this example is the direct analog of projecting a vector onto itself. 
The projection has the same magnitude as the vector, and the Cauchy--Schwarz inequality is saturated. 
The Fisher-information bound thus becomes an equality whenever the probe field is aligned with the score mismatch. 

This observation provides useful intuition for the general case. 
The Fisher information measures the total magnitude of the score mismatch, whereas a Schwinger--Dyson violation measures only the component visible to a particular probe. 
When the probe is perfectly aligned with the underlying probability distortion, the bound is saturated; when the probe is only partially aligned, the violation captures only part of the available information. 
This geometric interpretation will become central in our later discussion of the tomographic interpretation of the Schwinger--Dyson hierarchy. 

\subsection{Relative Entropy and Information Geometry} 

The appearance of Fisher information in the Schwinger--Dyson hierarchy reflects a deeper connection between Schwinger--Dyson violations and the geometry of probability distributions.

A central quantity in information theory is the relative entropy, or Kullback--Leibler divergence,
\begin{equation}
D_{\rm KL}(Q|P_{\rm eq}) = \left\langle \log \frac{Q}{P_{\rm eq}} \right\rangle_Q, 
\label{eq:KL} 
\end{equation}
which measures the distinguishability of the sampled distribution $Q$ from the equilibrium measure $P_{\rm eq}$. 
Unlike the score mismatch, which provides a point-wise description of the difference between the two distributions, the relative entropy is a global measure that compares them as complete probability measures. 

The relation between relative entropy and Fisher information plays a fundamental role in non-equilibrium statistical mechanics and information theory. 
For probability distributions $Q_t$, evolving under over-damped Fokker--Planck dynamics toward an equilibrium measure $P_{\rm eq}$, one has the entropy-production identity
\begin{equation}
\frac{d}{dt}
D_{\rm KL}(Q_t|P_{\rm eq}) = - I(Q_t|P_{\rm eq}),
\label{eq:entropy_production}
\end{equation}
under the standard assumptions of detailed balance and vanishing boundary contributions. 
This identity follows by differentiating the relative entropy, using the Fokker--Planck equation for $Q_t$, integrating by parts, and identifying the resulting quadratic form with the relative Fisher information~\cite{Stam1959}.

Equation~(\ref{eq:entropy_production}) shows that the relative Fisher information governs the instantaneous rate at which relative entropy is dissipated during relaxation toward equilibrium. 
Within the geometric framework developed here, Fisher information therefore admits a dual interpretation. 
It is both the squared norm of the score-mismatch field,
\begin{equation}
I(Q|P_{\rm eq}) = \langle |\delta s|^2 \rangle_Q,
\end{equation}
and, for Fokker--Planck evolution, the rate at which this probability-space distortion decays.

The geometric structure developed throughout this work is naturally expressed in terms of the inner product
\begin{equation}
(F,G)_Q \equiv \langle F \cdot G \rangle_Q, 
\label{eq:inner_product_FG} 
\end{equation} 
defined on square-integrable probe fields. 
In this notation, 
\begin{equation} 
I(Q|P_{\rm eq}) = (\delta s, \delta s)_Q, 
\end{equation} 
while 
\begin{equation} 
\Delta_F(Q) = - (F, \delta s)_Q. 
\end{equation} 

These relations summarize the geometric picture developed in this paper. 
The score mismatch is the fundamental geometric object characterizing departures from equilibrium. 
The relative Fisher information measures its norm, while each Schwinger--Dyson violation measures one of its projections onto a probe field. 
Relative entropy, Fisher information, and the Schwinger--Dyson hierarchy therefore provide complementary descriptions of the same underlying probability geometry: the relative entropy quantifies the global distinguishability of probability measures, the Fisher information measures the magnitude of the score mismatch, and the Schwinger--Dyson hierarchy resolves its directional structure. 

\section{Applications and Diagnostics} 
\label{sec:Applications_and_Diagnostics} 

The preceding sections established a geometric framework in which Schwinger--Dyson violations are identified with projections of the score-mismatch field, while the relative Fisher information measures its overall magnitude. We now turn to the practical implications of this picture. 

A primary motivation for studying Schwinger--Dyson identities is their role as diagnostics of sampling quality. 
In Monte Carlo simulations, stochastic quantization, and related stochastic algorithms, the sampled distribution generally differs from the target equilibrium measure because of incomplete equilibration, discretization effects, finite-volume corrections, or algorithmic approximations. 
Rather than simply signaling the presence of such errors, the geometric framework developed here provides a systematic way of characterizing their structure. 

We begin with configurational temperature, which provides perhaps the simplest and most familiar example of a Schwinger--Dyson diagnostic. 

\subsection{Configurational-Temperature Violations} 

Configurational temperature was originally introduced as an alternative estimator of temperature based entirely on derivatives of the action \cite{Rugh:1997, Rugh:1998}. 
Within the present framework, however, it acquires a broader interpretation. 
Rather than representing a distinguished observable, it appears as a particular member of the Schwinger--Dyson hierarchy corresponding to a specific choice of probe field. 

Recall that the configurational-temperature identity is obtained from the gradient vector field 
\begin{equation}
F = v = \frac{\nabla S_E}{|\nabla S_E|^2}, 
\label{eq:config_v} 
\end{equation} 
which satisfies $v \cdot \nabla S_E = 1$. 
The corresponding equilibrium Schwinger--Dyson identity is 
\begin{equation} 
\left\langle \nabla \cdot v \right\rangle_{\rm eq} = 1, 
\label{eq:config_equilibrium} 
\end{equation} 
motivating the definition 
\begin{equation} 
\beta_{\rm config}(Q) = \left\langle \nabla \cdot v \right\rangle_Q. 
\end{equation} 
When $Q = P_{\rm eq}$, one recovers $\beta_{\rm config} = 1$. 

Away from equilibrium, 
\begin{equation} 
\beta_{\rm config}(Q) - 1 = \left\langle \nabla \cdot v \right\rangle_Q - 1, 
\label{eq:config_deviation} 
\end{equation} 
and the general result of Sec.~\ref{sec:Non-Equilibrium_Probability_Geometry} immediately gives 
\begin{equation} 
\beta_{\rm config}(Q) - 1 = - \left\langle v \cdot \delta s \right\rangle_Q. 
\label{eq:config_projection} 
\end{equation} 

Equation~(\ref{eq:config_projection}) provides a direct geometric interpretation of configurational-temperature violations. 
The vector field $v$ defines a distinguished probe direction in configuration space, and the deviation of the configurational temperature from its equilibrium value is simply the projection of the score-mismatch field onto that direction. 
Configurational temperature therefore measures one specific component of the underlying probability distortion. 

The Fisher-information bound immediately yields 
\begin{equation} 
|\beta_{\rm config}(Q)-1|^2 \le \langle |v|^2 \rangle_Q \, I(Q|P_{\rm eq}), 
\label{eq:config_bound} 
\end{equation} 
showing that configurational-temperature violations are controlled by the overall magnitude of the score mismatch. 

This interpretation also clarifies the scope of configurational temperature as a diagnostic. 
Since it corresponds to only a single probe field, it is sensitive to only one projection of the score mismatch. 
Different probability distortions may therefore produce identical configurational-temperature violations, and some distortions may even remain invisible to this observable. 
A more complete characterization of sampling errors generally requires several independent Schwinger--Dyson probes that resolve complementary components of the underlying probability distortion. 

From this perspective, the significance of configurational temperature extends beyond its original role as a temperature estimator. 
It provides the simplest example of a much broader geometric principle: every Schwinger--Dyson identity acts as a directional probe of the score-mismatch field, and together these probes provide increasingly detailed information about departures from equilibrium. 

\subsection{Stochastic Quantization} 

The geometric framework developed above finds a natural application in stochastic quantization, where the probability distribution evolves continuously toward equilibrium in Langevin time. 
This setting provides a concrete realization of the probability geometry introduced in the preceding sections. 

Let $Q(\phi, \tau)$ denote the probability distribution at Langevin time $\tau$. 
Its evolution is governed by the Fokker--Planck equation associated with the Langevin process and, under suitable conditions, converges to the equilibrium measure 
\begin{equation} 
P_{\rm eq}(\phi) = \frac{1}{Z} e^{- S_E[\phi]}. 
\end{equation} 
The resulting dynamics may be viewed as a trajectory in the space of probability measures, an interpretation closely related to the geometric and variational formulations of Fokker--Planck evolution developed in the mathematical literature~\cite{JordanKinderlehrerOtto1998}. 

Within the present framework, equilibration is naturally characterized by the evolution of the score-mismatch field 
\begin{equation} 
\delta s(\tau) = \nabla \log \frac{Q(\phi, \tau)} {P_{\rm eq}(\phi)}, 
\label{eq:stochastic_delta_s} 
\end{equation} 
rather than by the probability distribution alone. 
At finite Langevin time, $\delta s(\tau)$ is generally nonzero, reflecting the deviation of the evolving distribution from equilibrium. 
As the system relaxes, the score mismatch decreases and eventually vanishes when $Q = P_{\rm eq}$. 

The corresponding Schwinger--Dyson violation associated with a probe field $F$ is 
\begin{equation} 
\Delta_F(\tau) = - \left\langle F \cdot \delta s(\tau) \right\rangle_{Q(\tau)}, 
\label{eq:stochastic_SD} 
\end{equation} 
so each member of the Schwinger--Dyson hierarchy monitors one component of the evolving score mismatch. 
At the same time, the relative Fisher information, 
\begin{equation} 
I(Q(\tau)|P_{\rm eq}) = \left\langle |\delta s(\tau)|^2 \right\rangle_{Q(\tau)}, 
\end{equation} 
provides a global measure of the remaining departure from equilibrium. Schwinger--Dyson violations and Fisher information therefore offer complementary descriptions of the relaxation process: the former resolve its directional structure, while the latter measures its overall magnitude. 

A simple illustration is provided by the Ornstein--Uhlenbeck process,
\begin{equation} 
\frac{dx}{d\tau} = - x + \eta(\tau), 
\label{eq:OU_langevin} 
\end{equation} 
where 
\[ 
\langle \eta(\tau) \rangle = 0, \qquad \langle \eta(\tau) \eta(\tau') \rangle = 2 \delta(\tau - \tau'). 
\] 
The equilibrium distribution is 
\begin{equation} 
P_{\rm eq}(x) = \frac{1}{\sqrt{2 \pi}} e^{- x^2 / 2}, 
\label{eq:OU_equilibrium} 
\end{equation} 
corresponding to the action 
\[ 
S_E(x) = \frac12 x^2. 
\] 

Suppose the system is initialized in the displaced Gaussian 
\begin{equation} 
Q(x, \tau) = \frac{1}{\sqrt{2 \pi}} \exp \left[ - \frac{(x - \mu(\tau))^2}{2} \right], 
\label{eq:OU_distribution} 
\end{equation} 
whose mean evolves according to 
\begin{equation} 
\mu(\tau) = \mu_0 e^{-\tau}. 
\label{eq:OU_mean} 
\end{equation} 

The equilibrium score field is 
\[ 
s_{\rm eq} = - x, 
\] 
while the evolving distribution has 
\[ 
s_Q = - (x - \mu(\tau)). 
\] 
Hence, 
\begin{equation} 
\delta s(\tau) = \mu(\tau) = \mu_0 e^{-\tau}, 
\label{eq:OU_score} 
\end{equation} 
showing that the entire departure from equilibrium is encoded in a single exponentially decaying parameter. 

The corresponding Schwinger--Dyson violation is 
\begin{equation} 
\Delta_F(\tau) = - \mu_0 e^{- \tau} \langle F \rangle_{Q(\tau)}, 
\label{eq:OU_SD} 
\end{equation} 
so every Schwinger--Dyson probe inherits the exponential relaxation of the score mismatch. 
The detailed time dependence depends on the chosen probe field through $\langle F\rangle_{Q(\tau)}$, but the common driving mechanism is the decay of the underlying score-mismatch field. 

Although the Ornstein--Uhlenbeck process is exceptionally simple, it illustrates the general geometric picture. 
In more complicated systems the score mismatch typically develops a nontrivial spatial structure, and different Schwinger--Dyson probes may relax at different rates. 
Nevertheless, the underlying interpretation remains unchanged: equilibration corresponds to the disappearance of the score mismatch field, while the Schwinger--Dyson hierarchy provides a family of complementary probes that monitor this process throughout the evolution. 

\subsection{Complex Langevin Dynamics} 

An interesting direction in which the present framework may prove useful is complex Langevin dynamics. 
In theories with complex actions, conventional importance sampling is obstructed because the Boltzmann weight is no longer positive definite. 
Complex Langevin circumvents this difficulty by complexifying the configuration space and evolving the resulting variables according to a stochastic process. 
Although the method has been successfully applied to many systems, convergence of the Langevin process does not, by itself, guarantee convergence to the correct expectation values. 
Understanding and diagnosing such failures remains an important problem in applications ranging from finite-density field theory to matrix models. 

A central feature of the modern analysis of complex Langevin dynamics is the role played by Schwinger--Dyson identities. 
The formal justification of the method relies on integration-by-parts arguments relating expectation values computed with the stationary Langevin distribution to the Schwinger--Dyson hierarchy of the underlying theory. 
When these arguments fail, for example because of insufficient localization in the complexified configuration space or the presence of boundary contributions, the stationary distribution need not satisfy the desired Schwinger--Dyson identities, leading to incorrect convergence. 
These ideas were first systematically developed by Guralnik and Pehlevan~\cite{Pehlevan:2007eq} and have recently been refined into increasingly sharp correctness criteria based on the recovery of the Schwinger--Dyson hierarchy and the validity of the underlying integration-by-parts relations~\cite{Mandl:2025mav,Mandl:2026vdc}.

The geometric framework developed in this work is closely related in spirit to these developments, but addresses a different problem. 
Throughout this paper we have assumed the existence of a real equilibrium probability measure, $P_{\rm eq}$, allowing Schwinger--Dyson violations to be expressed in terms of the score-mismatch field
\[
\delta s = \nabla \log \frac{Q}{P_{\rm eq}}.
\] 
This construction does not directly extend to complex Langevin dynamics, where the target weight is generally complex and cannot itself be interpreted as a probability distribution. 

Nevertheless, the central role of Schwinger--Dyson identities in both settings suggests that a similar geometric description may exist for complexified stochastic dynamics. 
An interesting open question is whether one can formulate an analogue of the score-mismatch field that characterizes departures from correct convergence in the complexified configuration space. 
Such a construction could provide a geometric interpretation of incorrect convergence and organize different Schwinger--Dyson diagnostics according to the components of the underlying probability distortion that they probe. 

Developing such a framework lies beyond the scope of the present work. 
The geometric interpretation introduced here nevertheless suggests a possible route toward extending score-based descriptions of probability geometry beyond real equilibrium measures. 

\subsection{Practical Implementation} 

The geometric interpretation developed above naturally suggests a practical strategy for diagnosing equilibration in stochastic simulations. 
Rather than relying on a small collection of observables, one may monitor a family of Schwinger--Dyson identities associated with different probe fields. 
Since each probe is sensitive to a different component of the score mismatch, the resulting diagnostics provide complementary information about departures from equilibrium. 

Given a collection of probe fields
\[
F_1, F_2, \ldots, F_n, 
\] 
one computes the corresponding violations
\[ 
\Delta_{F_1}, \Delta_{F_2}, \ldots, \Delta_{F_n}. 
\] 
Agreement among several independent probes provides considerably stronger evidence for equilibration than any individual diagnostic in isolation. 
Conversely, discrepancies between different probes can reveal anisotropic or localized probability distortions that would remain invisible to a single observable. 

Within this family, configurational temperature occupies a distinguished position because it corresponds to the probe 
\[ 
v = \frac{\nabla S_E}{|\nabla S_E|^2}. 
\] 
Its geometric interpretation makes it a natural first diagnostic to examine. 
At the same time, the present framework clarifies its limitations: as a single projection of the score mismatch, configurational temperature cannot by itself detect every departure from equilibrium. 
Different probability distortions may produce identical configurational-temperature violations or even leave the configurational temperature unchanged. 

The computational cost of Schwinger--Dyson diagnostics depends on the chosen probe fields. For configurational temperature, one must evaluate 
\begin{equation} 
\nabla \cdot \left( \frac{\nabla S_E}{|\nabla S_E|^2} \right), 
\label{eq:config_cost} 
\end{equation}
which involves both first and second derivatives of the action. 
In many lattice and matrix-model simulations the gradient of the action is already available as part of the update algorithm, so the principal additional cost arises from computing its divergence. 
Other probe fields may be substantially cheaper or more expensive, allowing the diagnostic strategy to be adapted to the computational demands of a particular problem. 

More broadly, the geometric framework developed here suggests that Schwinger--Dyson identities should not be viewed merely as pass--fail tests of equilibration. 
Each violation contains quantitative information about a particular component of the underlying probability distortion. 
By selecting appropriate families of probe fields, one can therefore obtain a progressively richer picture of how the sampled distribution differs from the target equilibrium measure. 

These observations motivate the next section. Since different Schwinger--Dyson identities probe different components of the score mismatch field, it is natural to ask how much information about the underlying probability distortion is encoded in the hierarchy as a whole. 
This question leads directly to a tomographic interpretation of Schwinger--Dyson violations. 


\section{Tomography of Probability Geometry} 
\label{sec:Tomography_of_Probability_Geometry} 

The geometric framework developed in the preceding sections identifies every Schwinger--Dyson violation as a projection of the score-mismatch field, 
\begin{equation} 
\Delta_F(Q) = -(F, \delta s)_Q. 
\end{equation} 
This observation naturally raises the question of how much information about the underlying probability distortion is encoded in the Schwinger--Dyson hierarchy as a whole. 

The idea that the Schwinger--Dyson hierarchy contains detailed information about the underlying probability measure has appeared previously in the analysis of complex Langevin dynamics~\cite{Pehlevan:2007eq}. 
The perspective developed here is complementary. Rather than viewing Schwinger--Dyson identities primarily as consistency conditions, we interpret them as geometric measurements of the score-mismatch field. 
This viewpoint suggests a tomographic interpretation in which different probe fields reveal complementary aspects of the underlying probability distortion. 

\subsection{Can the Schwinger--Dyson Hierarchy Determine the Fisher Information?} 

Section~\ref{sec:Fisher_Information_and_SD_Violations} established that small Fisher information implies small Schwinger--Dyson violations for every probe field. 
A natural question is whether some converse statement also holds. 
Can the information contained in the complete Schwinger--Dyson hierarchy be used to determine the magnitude of the score mismatch itself? 

For an individual probe field the answer is clearly negative. 
If 
\begin{equation}  
\Delta_F(Q) = 0, 
\end{equation} 
then 
\begin{equation}
(F, \delta s)_Q = 0, 
\end{equation} 
which constrains only a single projection of the score-mismatch field. 
Components orthogonal to the chosen probe remain completely unconstrained. 
This is entirely analogous to ordinary Euclidean geometry: knowing that the projection of a vector onto one direction vanishes does not imply that the vector itself vanishes. 

The situation changes once one considers the complete family of probe fields. 
Using the inner product $(F, G)_Q = \langle F \cdot G \rangle_Q$, the Schwinger--Dyson violation defines a bounded linear functional on the Hilbert space of square-integrable probe fields, $\Delta_F(Q) = - (F, \delta s)_Q$. 
Applying the Cauchy--Schwarz inequality gives 
\begin{equation} 
|\Delta_F(Q)|^2 \le (F, F)_Q \, (\delta s, \delta s)_Q. 
\end{equation} 
Since 
\begin{equation} 
(\delta s, \delta s)_Q = I(Q | P_{\rm eq}), 
\end{equation} 
one immediately obtains 
\begin{equation} 
\frac{|\Delta_F(Q)|^2} {\langle |F|^2\rangle_Q} \le I(Q | P_{\rm eq}), 
\label{eq:tomo_bound} 
\end{equation} 
for every admissible probe field. 

The inequality is saturated when the probe field is proportional to the score mismatch, 
\begin{equation} 
F \propto \delta s, 
\end{equation} 
so that 
\begin{equation}
I(Q|P_{\rm eq}) = \sup_{F \in L^2(Q)} \frac{|\Delta_F(Q)|^2} {\langle |F|^2\rangle_Q}.
\label{eq:FisherVariational}
\end{equation}
The supremum is taken over square-integrable probe fields with respect to the inner product $(F, G)_Q$. 

Equation~(\ref{eq:FisherVariational}) provides a variational characterization of the relative Fisher information entirely in terms of Schwinger--Dyson violations. 
Rather than being introduced independently, the Fisher information may be interpreted as the largest normalized Schwinger--Dyson violation attainable within the complete hierarchy of probe fields. 

This result has an important conceptual consequence. Although an individual Schwinger--Dyson identity contains only partial information about the score mismatch, the hierarchy as a whole encodes its norm. The Schwinger--Dyson hierarchy therefore contains considerably more information than is apparent from any single probe and naturally suggests a tomographic interpretation of probability-space distortions. We now explore this interpretation in more detail. 

\subsection{A Tomographic Interpretation of Sampling Errors} 

Equation~(\ref{eq:FisherVariational}) suggests a natural geometric interpretation of the Schwinger--Dyson hierarchy. 
Although an individual Schwinger--Dyson identity probes only one projection of the score-mismatch field, the complete hierarchy contains information about the score mismatch as a whole. 
This observation motivates viewing Schwinger--Dyson diagnostics through the lens of tomography. 

The analogy is straightforward. In tomographic imaging an unknown object is not observed directly; instead, one measures a collection of projections along different directions and reconstructs progressively more information about the underlying structure. 
No individual projection is sufficient, but a sufficiently rich family of projections reveals the object with increasing fidelity. 

The Schwinger--Dyson hierarchy admits a closely analogous interpretation. 
The score-mismatch field is generally not directly accessible in a stochastic simulation. Instead, one measures the Schwinger--Dyson violations $\Delta_{F_1}, \Delta_{F_2}, \ldots, \Delta_{F_n}$, associated with a family of probe fields $F_1, F_2, \ldots, F_n$. 
Each violation measures a single projection of the score mismatch, 
\begin{equation} 
\Delta_{F_i}(Q) = - (F_i, \delta s)_Q, 
\end{equation} 
so different probe fields reveal complementary information about the underlying probability distortion.

This viewpoint changes the role of Schwinger--Dyson identities. 
Rather than serving only as consistency conditions whose violation signals imperfect sampling, they may also be regarded as quantitative probes of the geometry of the sampling error. 
Individual probes resolve different directional components of the score mismatch, while increasingly rich families of probe fields provide progressively more detailed information about its structure. 

Within this picture, the relative Fisher information plays a complementary role. 
Whereas the Schwinger--Dyson hierarchy resolves the directional structure of the score mismatch, the Fisher information, $I(Q|P_{\rm eq}) = \langle |\delta s|^2 \rangle_Q$, measures its overall magnitude. 
The hierarchy and the Fisher information therefore provide complementary descriptions of the same probability-space distortion: one resolves its directional content, while the other quantifies its total size. 

In practice one rarely has access to the complete hierarchy of probe fields. 
Nevertheless, even a modest collection of independent Schwinger--Dyson identities can provide substantially more information than any single diagnostic considered in isolation. 
The geometric framework developed here therefore suggests that convergence diagnostics should be viewed not merely as pass-fail tests, but as measurements that characterize the structure of residual sampling errors. 

\subsection{Example: Gaussian Sampling-Error Tomography} 

The tomographic interpretation becomes particularly transparent for multidimensional Gaussian distributions, where all quantities can be evaluated analytically. 

Consider the equilibrium distribution 
\begin{equation} 
P_{\rm eq}(x) = \frac{(\det A)^{1/2}} {(2 \pi)^{n/2}} \exp \left( - \frac12 x^T A x \right), 
\label{eq:gaussian_tomo_peq} 
\end{equation} 
where $A$ is a symmetric positive-definite matrix. 
Suppose that the sampled distribution is 
\begin{equation} 
Q(x) = \frac{(\det B)^{1/2}} {(2 \pi)^{n/2}} \exp \left( - \frac12 x^T B x \right), 
\label{eq:gaussian_tomo_Q} 
\end{equation} 
with $B$ also symmetric and positive definite. 

The corresponding score fields are 
\begin{equation} 
s_{\rm eq} = - A x, \qquad s_Q = - B x, 
\end{equation} 
so the score mismatch is 
\begin{equation} 
\delta s = -(B - A) x. 
\end{equation} 

Introducing the distortion matrix 
\begin{equation} 
D = B - A, 
\end{equation} 
one obtains 
\begin{equation} 
\delta s = - D x. 
\label{eq:gaussian_delta_s} 
\end{equation} 
The entire departure from equilibrium is therefore encoded in the linear map $D$. 

To probe this distortion, consider the family of vector fields 
\begin{equation} 
F_u(x) = u \, (u^T x), 
\end{equation} 
where $u$ is an arbitrary constant vector. 
These probe fields measure fluctuations along the direction $u$. 

Using 
\[ 
\delta s = - D x, 
\] 
the Schwinger--Dyson violation becomes 
\begin{align} 
\Delta_u &= - (F_u, \delta s)_Q \nonumber \\ 
&= \left\langle (u^T D x) (u^T x) \right\rangle_Q \nonumber \\ 
&= u^T D B^{-1} u. 
\label{eq:gaussian_projection} 
\end{align} 

Since a quadratic form depends only on the symmetric part of its matrix, 
\begin{equation} 
u^T D B^{-1} u = u^T \left( \frac{D B^{-1} + B^{-1} D}{2} \right) u, 
\end{equation} 
the Schwinger--Dyson hierarchy probes the symmetric part of the distortion operator 
\begin{equation} 
M = \frac12 \left( D B^{-1} + B^{-1} D \right). 
\end{equation} 

Different probe directions therefore reveal different aspects of the sampling error. 
Each choice of $u$ determines one quadratic form of the symmetric operator $M$, while increasingly rich families of probe directions provide progressively more information about its structure. 
In this Gaussian setting, the Schwinger--Dyson hierarchy therefore admits a clear tomographic interpretation. 

The corresponding relative Fisher information is 
\begin{equation} 
I(Q|P_{\rm eq}) = {\rm Tr} \left( D^2 B^{-1} \right), 
\end{equation} 
which measures the overall magnitude of the covariance distortion. 
The Fisher information therefore provides a global characterization of the departure from equilibrium, while the Schwinger--Dyson hierarchy resolves its directional structure through the family of quadratic forms (\ref{eq:gaussian_projection}). 

If the probe directions span $\mathbb{R}^n$ sufficiently densely, the associated quadratic forms determine the symmetric operator $M$. 
This illustrates, in a simple finite-dimensional setting, how families of Schwinger--Dyson probes can provide progressively richer information about the underlying probability distortion without requiring direct access to the score-mismatch field itself. 

\subsection{Selecting Effective Probe Fields} 

The tomographic interpretation naturally raises a practical question: how informative is a given probe field? 
Since different Schwinger--Dyson identities measure different projections of the score mismatch, some probe fields will be considerably more sensitive to a particular sampling error than others. 

The answer follows directly from the Cauchy--Schwarz inequality. Writing the probe field as 
\begin{equation} 
F = F_{\parallel} + F_{\perp}, 
\end{equation} 
where $F_{\parallel}$ and $F_{\perp}$ denote the components parallel and orthogonal to the score-mismatch field, respectively, one has 
\begin{equation} 
\Delta_F(Q) = - (F, \delta s)_Q = - (F_{\parallel}, \delta s)_Q. 
\end{equation} 
The orthogonal component therefore makes no contribution to the Schwinger--Dyson violation. 

This observation motivates the dimensionless quantity 
\begin{equation} 
R_F = \frac{|\Delta_F(Q)|^2} {\langle |F|^2\rangle_Q \, I(Q|P_{\rm eq})}, 
\label{eq:sharpness_ratio} 
\end{equation} 
which satisfies 
\begin{equation} 
0 \le R_F \le 1. 
\end{equation} 

Using the orthogonal decomposition of $F$, one finds 
\begin{equation} 
R_F = \frac{(F_{\parallel}, F_{\parallel})_Q}{(F_{\parallel}, F_{\parallel})_Q + (F_{\perp}, F_{\perp})_Q}. 
\label{eq:sharpness_geometry} 
\end{equation} 
Equivalently, 
\begin{equation} 
R_F = \cos^2 \theta, 
\end{equation} 
where $\theta$ is the angle between the probe field and the score-mismatch field in the Hilbert space defined by the inner product $(F, G)_Q$. 

The geometric interpretation is immediate. 
Probe fields aligned with the score mismatch satisfy $R_F \approx 1$ and are highly sensitive to the underlying sampling error. 
Conversely, probe fields nearly orthogonal to the score mismatch satisfy $R_F\approx0$ and detect little or no violation, even when the sampled distribution differs significantly from equilibrium. 

In practice, the score-mismatch field is unknown and therefore $R_F$ cannot be evaluated directly. 
Nevertheless, this result provides useful guidance for constructing effective diagnostics. 
Different families of probe fields emphasize different aspects of the probability distortion, and combining several independent probes generally provides a more robust assessment of equilibration than relying on any single observable. 

\section{Discussion and Outlook} 
\label{sec:Discussion_and_Outlook} 

The traditional role of Schwinger--Dyson identities is to characterize equilibrium probability measures through an infinite hierarchy of exact constraints. 
In this work we have proposed a complementary perspective. 
Rather than viewing Schwinger--Dyson identities solely as consistency conditions, we have argued that they may also be regarded as geometric measurements of probability distributions. 

The central object underlying this interpretation is the score-mismatch field, 
\begin{equation} 
\delta s = \nabla \log \frac{Q}{P_{\rm eq}}, 
\end{equation} 
which compares the score of an arbitrary sampled distribution $Q$ with that of the equilibrium measure $P_{\rm eq}$. 
Within this framework, departures from equilibrium are no longer described simply by the failure of individual identities, but by the emergence of a geometric vector field on configuration space. 
Every Schwinger--Dyson violation measures a projection of this field, 
\begin{equation} 
\Delta_F(Q) = - (F, \delta s)_Q, 
\end{equation} 
while the relative Fisher information, 
\begin{equation} 
I(Q|P_{\rm eq}) = (\delta s, \delta s)_Q, 
\end{equation} 
measures its squared norm. 
The Schwinger--Dyson hierarchy and the Fisher information therefore appear as complementary manifestations of the same underlying geometric structure. 

Viewed from this perspective, configurational temperature acquires a natural interpretation as one particular member of a much larger family of geometric probes. 
Likewise, the Fisher-information bound and the variational characterization derived in this work show that the Schwinger--Dyson hierarchy contains considerably richer information than is apparent from any individual identity. 
This naturally leads to a tomographic interpretation in which different probe fields reveal complementary directional components of the underlying score mismatch. 
Although a complete reconstruction of the score-mismatch field generally requires infinitely many probes, even finite families of Schwinger--Dyson identities provide substantially more information about sampling errors than any single diagnostic considered in isolation. 

One of the appealing features of this formulation is that it brings together several ideas that have largely developed independently. 
Configurational temperature, Schwinger--Dyson identities, relative Fisher information, Stein operators, and score-function methods all emerge naturally from the same probability-geometric framework. 
Rather than introducing new diagnostic quantities, the present work identifies the geometric object from which these seemingly disparate constructions arise. 
In this sense, the score-mismatch field provides a common language for describing departures from equilibrium across a broad range of stochastic sampling methods. 

The geometric viewpoint developed here also suggests several directions for future investigation. 
One natural question is whether the score-mismatch field can be approximately reconstructed from finite collections of Schwinger--Dyson measurements, leading to practical tomographic algorithms for diagnosing sampling errors in Monte Carlo simulations and stochastic quantization. 
A second direction is the extension of these ideas to interacting lattice field theories and matrix models, where different probe fields may become sensitive to physically distinct classes of probability distortions. 
Finally, it would be interesting to investigate whether an analogue of the score-mismatch field can be formulated for complexified probability spaces, thereby extending the present framework to complex Langevin dynamics, where Schwinger--Dyson identities already play a central role in modern correctness criteria. 

More broadly, we believe that the geometric viewpoint developed here shifts the emphasis from individual observables to the structure of the underlying probability distribution itself. 
Rather than asking only whether a simulation has converged, one may ask how the sampled distribution differs from the target measure and which components of that difference are detected by different observables. 
The score-mismatch field provides a natural language for addressing these questions. 

The principal message of this work may therefore be summarized succinctly. Schwinger--Dyson identities do more than characterize equilibrium probability measures. 
They also provide geometric probes of probability distributions. 
Viewed through the score-mismatch field, the Schwinger--Dyson hierarchy becomes a family of measurements of probability geometry, while the Fisher information measures the magnitude of the underlying distortion. 
Together they offer a unified geometric framework for understanding non-equilibrium sampling, equilibration, and stochastic dynamics. 

\acknowledgments

We extend our gratitude to Sumit Das, Navdeep Singh Dhindsa, Vamika Longia, Michael Mandl, and Piyush Kumar for their invaluable discussions. 
The work of A.J. was supported in part by a Start-up Research Grant from the University of the Witwatersrand. 

\raggedright
\bibliographystyle{utphys}
\bibliography{bibfile}
\end{document}